\date{}
\documentclass[aps,pra,showpacs,twocolumn]{revtex4}
\usepackage{graphicx,psfrag,amsmath,amssymb,amsfonts,latexsym,color,dcolumn}

\begin{document}

\title{Environmentally-induced effects on a bipartite two-level system: geometric phase and entanglement
properties}

\author{Fernando C. Lombardo$^1$ \footnote{lombardo@df.uba.ar}}
\author{Paula I. Villar$^{1,2}$ \footnote{paula@df.uba.ar}}

 \affiliation{$^1$ Departamento de F\'\i sica {\it Juan Jos\'e
 Giambiagi}, FCEyN UBA, Facultad de Ciencias Exactas y Naturales,
 Ciudad Universitaria, Pabell\' on I, 1428 Buenos Aires, Argentina \\
$^2$ Computer Applications on Science and Engineering Department,
Barcelona Supercomputing Center (BSC),
29, Jordi Girona
08034 Barcelona,
Spain}

\date{today}

\begin{abstract}
We calculate the geometric phase of a bipartite two-level system
coupled to an external environment. We analyze the reduced density
matrix for an arbitrary initial state of the composite system and
compute the correction to the unitary geometric phase through a
kinematic approach. In all cases considered, we observe a similar structure
 as a function of the degree of the entanglement of the initial state.
Further, we compute the entanglement entropy and
concurrence of the bipartite state and analyze if there is any
relation among these quantities and the geometric phase acquired
during the nonunitary system's evolution. Finally, we discuss the
results obtained.
\end{abstract}

\pacs{03.65.Vf,03.65.Ud,03.67.Pp}

\maketitle

\newcommand{\beq}{\begin{equation}}
\newcommand{\eeq}{\end{equation}}
\newcommand{\beqa}{\begin{eqnarray}}
\newcommand{\eeqa}{\end{eqnarray}}
\newcommand{\beqas}{\begin{eqnarray*}}
\newcommand{\eeqas}{\end{eqnarray*}}

\section{Introduction}

It is plausible to imagine a quantum computer as a device made of
qubits which interact among themselves in some way. Therefore, a
quantum computer can be understood as an interacting quantum many
body system. In the last years, several results have been obtained
which suggest that entanglement is at the root of the power of
quantum computers \cite{Vidal}.

A qubit is a quantum two-level system, i.e. a physical system
described in terms of a Hilbert space $C^2$. This can be considered
as a spin-1/2 particle, or as an atom with two energy levels or even
a photon with two possible orthogonal polarizations. Solid state
examples which have achieved some experimental success include spins
in semiconductors and quantum dots, various designs based on
superconductors, vacancy centers in diamond and singles molecule
magnets (SMMs), see \cite{Stamp} and references therein.

It is has been shown that one can describe a quantum information
processing system in terms of interacting qubits such as SMMs. Quite
generally, in doing so, we need to include both environmental
nonlocalized modes (like phonons and photons) and discrete localized
modes (like defects, nuclear spins, loose spins). The spin-boson
model corresponds to a single two-level system interacting with a
large reservoir of bosonic field modes, i.e. a spin-$1/2$ particle
coupled to an environment of harmonic oscillators. The seminal
review paper by Legget et al. \cite{Leggett} discusses its dynamics
in great detail. The spin-spin model is its analogous but when the
spin-$1/2$ particle is coupled to an environment of spin-$1/2$
particles. The effect of this type of environment on the central
spin has also been studied thoroughly by Zurek in Ref.\cite{Zurek}.

When coherent superposition
states on macroscopic scales are built, they are tipically created at
very low temperatures in order to ``freeze out'' the thermal
environment and thus minimize the decoherence effects. This thermal
environment can be modelled into a bath of delocalized
bosonic modes, implying an interaction among a large
spatial region. But lowering the temperature does not affect
the influence of localized modes such as nuclear spins or
impurities that are intrinsically present in the material
\cite{schlosshauer}.

From another point of view, a system can retain the information of
its motion when it undergoes a cyclic evolution, in the form of a
geometric phase (GP), which was first put forward by Pancharatnam in
optics \cite{Pancharatman} and later studied explicitly by Berry in
a general quantal system \cite{Berry}. Since then, great progress
has been achieved in this field.  The geometric phase has been
extended to the case of non-adiabatic evolutions \cite{Anandan}. As
an important evolvement, the application of the geometric phase has
been proposed in many fields, such as the geometric quantum
computation. Due to its global properties, the geometric phase is
propitious to  construct  fault tolerant  quantum  gates. In this
line of work, many physical systems  have  been  investigated  to
realize  geometric  quantum  computation,  such  as  NMR  (Nuclear
Magnetic Resonance), Josephson  junction,  Ion  trap  and
semiconductor  quantum  dots. The quantum computation scheme for the
geometric phase has been proposed based on the Abelian  or
non-Abelian geometric phase, in which geometric phase has been shown
to be robust against faults in the presence of some kind of
external noise due to the geometric nature of Berry phase. It was
therefore seen that interactions play an important role for the
realization of some specific operations. Consequently, the study of the
geometric phase was soon extended to open quantum systems. Following this idea,
 many authors have analyzed the correction to the geometric
phase under the influence of an external environment using different
approaches \cite{Whitney, Carollo, dechiara, Tong, PRA, nos, pau}.

In this context, we shall briefly review  the way the
geometric phase can be computed for a system under
the influence of external conditions such as an external bath.
 In Ref. \cite{Tong}, a quantum kinematic
approach was proposed and the geometric phase
(GP) for a mixed state
under nonunitary evolution has been defined
as
\begin{eqnarray} \phi_G & = &
{\rm arg}\bigg\{\sum_k \sqrt{ \varepsilon_k (0) \varepsilon_k (\tau)}
\langle\Psi_k(0)|\Psi_k(\tau)\rangle \times \nonumber \\
& & e^{-\int_0^{\tau} dt \langle\Psi_k|
\frac{\partial}{\partial t}| {\Psi_k}\rangle}\bigg\}, \label{fasegeo}
\end{eqnarray}
where $\varepsilon_k(t)$ are the eigenvalues and
 $|\Psi_k\rangle$ the eigenstates of the reduced density matrix
$\rho_{\rm r}$ (obtained after tracing over the reservoir degrees
 of freedom). In the last definition, $\tau$ denotes a time
after the total system completes a cyclic evolution when it is
isolated from the environment. Taking into account the effect of the
environment, the system no longer undergoes a cyclic evolution.
However, we shall consider a quasi cyclic path ${\cal P}:~ t~
\epsilon~ [0,\tau]$, with $\tau= 2 \pi/ \Omega$ ($\Omega$ is the
system's characteristic frequency). When the system is open, the
original GP, i.e. the one that would have been obtained if the
system had been closed $\phi_G^U$, is modified. This means, in a
general case, the phase is $\phi_G = \phi^U_G + \delta \phi_G$,
where $\delta \phi_G$ depends on the kind of environment coupled to
the main system \cite{PRA,nos,pau,Benarjee, Rezakhani}. If the
eigenvalues of the density matrix are degenerate, the expression for
the geometric phase takes a slightly different form, as described in
\cite{Tong}.

In this paper, we shall study the geometric phase acquired by a
bipartite system in the presence of an external environment. We
shall consider both the presence of a bosonic and spin environment.
We shall choose an arbitrary initial state and see how  the
geometric phase acquired during the evolution of the composite
system is corrected by the presence of an environment. We will show
the dependence of the environmentally induced correction to the
geometric phase upon the degree of entanglement in the bipartite
system. We will also show that, even in the nonunitary evolution,
there is no correction to the phase when the composite system is in
a maximally entangled state (MES). In this case, we show that the
system adopt a total phase $\phi_G^U = \pi$ which, as it has been
shown in previous articles, is of topological nature \cite{milman}.
Furthermore, we shall also study the entanglement measures of the
bipartite system and analyze the relation among them and the
geometric phase. 

This paper is organized as follows. In Section II,
we study the evolution, geometric phase, and entanglement properties
of the bipartite system in the presence of a bosonic environment at
zero temperature. In Section III, we follow the same procedure for a
spin environment leaving its trail in the dynamics of the bipartite
system. Finally, we conclude our results in Section IV. Two appendices  
complete the presentation.

\section{Bipartite Spin-Boson Model}

Oscillator environments correspond to a quasi continuum
of delocalized bosonic field modes, with coherence
and energy from the central system becoming effectively
and irreversibly lost into this extended bosonic environment.

We shall consider a bipartite system, that is to say, two
interacting two-level systems, both coupled to an external
reservoir.

\begin{figure}[h!t]
\begin{center}
\includegraphics[width=6cm]{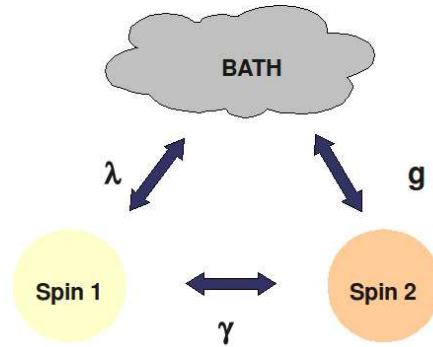}
 \caption{Diagram of the model we shall consider: two two-level systems
coupled to an external reservoir whether it is bosonic or comprised of
spins.}
\label{modelito}
\end{center}
 \end{figure}

%\end{minipage}
In terms of the Hamiltonians, the model can be mathematically
described by the Hamiltonian of the free bipartite system $H_S$, the
Hamiltonian of interaction between the bipartite and the external
bath $H_I$ and the free Hamiltonian of the external bath $H_B$:
\begin{eqnarray}
H_S &=& \frac{\hbar \Omega_1}{2} \sigma_z^1 + \frac{\hbar
\Omega_2}{2} \sigma_z^2 + \gamma~ \sigma_z^1 \otimes \sigma_z^2\\
H_I &=& \sigma_z^1 \otimes \sum_{n=1}^N \lambda_n q_n +
\sigma_z^2 \otimes \sum_{n=1}^N g_n q_n \\
H_B &=& \sum_{n=1}^N \hbar \omega_n a_n^{\dagger} a_n,
\end{eqnarray}
where the constants $\lambda_n$ and $g_n$ couple the system to each
oscillator in the environment, and $\gamma$ is the coupling strength
between both spin-$1/2$ particles (as shown in Fig.\ref{modelito}).
Here we have assumed that each coupling constant of the two level systems
with the environment is different being $\lambda_n$ for the spin 1 
and $g_n$ for spin 2.

In order to compute the geometric phase for the bipartite system it
is important to know its dynamics at all times. Therefore, in 
Appendix \ref{derivationSB} we have derivated the reduced density matrix
of the bipartite system. It is important to know that this derivation
has been done in the weak coupling limit for a general environment
defined by a spectral function $J(\omega)$. Following the mentioned
derivation, if one consideres the most general case for an initial state
of the bipartite system, namely
\begin{equation}
| \Phi(0) \rangle = \alpha |0 0 \rangle + \beta | 0 1 \rangle  + \zeta
|1 0 \rangle + \delta | 1 1 \rangle,
\label{estadogen}
\end{equation}
the reduced density matrix
for this model can be written as,

\begin{widetext}
\begin{equation}
\rho_{\rm r}(t)=\left(\begin{array}{cccc} |\alpha|^2 & \alpha
\beta^* e^{-i(2 \gamma +\Omega_2)t} \Gamma_{2} \varLambda_{12} &
\alpha \zeta^*  e^{-i(2 \gamma +\Omega_1)t} \Gamma_{1}
\varLambda_{12} & \alpha \delta^*
e^{-i(\Omega_1+\Omega_2)t} \Gamma_1 \Gamma_2 \Gamma_{12}^2 \\
\beta \alpha^*  e^{i(2 \gamma +\Omega_2)t} \Gamma_{2}
\varLambda_{12}^*  &
 |\beta|^2  & \beta \zeta^*
e^{-i(\Omega_1-\Omega_2)t} \Gamma_1 \Gamma_2 {\tilde \Gamma}_{12}^2 &
\beta \delta^* e^{-i(\Omega_1 -2 \gamma)t} \Gamma_1 \varLambda_{12}^*\\
\zeta \alpha^* e^{i(2 \gamma +\Omega_1)t} \Gamma_{1}
\varLambda_{12}^* & \zeta \beta^* e^{i(\Omega_1-\Omega_2)t} \Gamma_1
\Gamma_2  {\tilde \Gamma}_{12}^2 & |\zeta|^2  &
 \zeta \delta^* e^{-i(\Omega_2-2 \gamma)t} \Gamma_2 \varLambda_{12}^* \\
\delta \alpha^* e^{i(\Omega_1+\Omega_2)t} \Gamma_1 \Gamma_2
\Gamma_{12}^2 & \delta \beta^* e^{i(\Omega_1 -2 \gamma)t} \Gamma_1
\varLambda_{12} & \delta \zeta^* e^{i(\Omega_2-2 \gamma)t} \Gamma_2
\varLambda_{12}  & |\delta|^2
\end{array}\right), \nonumber
\end{equation}
\end{widetext}
where we have not written explicitly the dependence upon the time of
$\Gamma_i(t)$ just to simplify notation and we have defined
\begin{equation}
{\tilde \Gamma}_{12}^2={\tilde \Gamma}_{12}^2(t)=e^{+ 8 \int_0^t dt_1
F_{12}(t_1)}
\end{equation}
and
\begin{equation}
\varLambda_{12}=\varLambda_{12}(t)=e^{{\rm i}~ 4 \int_0^t dt_1
G_{12}(t_1)}
\end{equation}
with $\Gamma_1$, $\Gamma_2$, $F_i(t)$ and $G_{i}(t)$ as defined in 
Appendix \ref{derivationSB}. Notice
that the coupling constants of the model $\lambda_n$
and $g_n$ are absorbed in the dimensionless dissipative
constants ${\gamma_0}_1 \sim \lambda^2$, ${\gamma_0}_2 \sim g^2$, and ${\gamma_0}_{12}\sim \lambda g$
respectively, defined in the spectral density of the bath (see Appendix \ref{derivationSB} 
for details).

In order to have a complete description of the effect of the bath on
the bipartite system, we shall consider a quantum ohmic environment
at zero temperature and different couplings between both spin-particles 
and the bath. Consequently, there will be different
decoherence factors (see Appendix \ref{derivationSB}) and we shall use 
subindexes 1,2, or 12 to refer to them. By the definition of the particular initial state, the
reduced density matrix and consequently, the dynamics of the open
bipartite system is known.

For the particular purpose of this work, we shall define an initial
density matrix of the form (by setting the corresponding values of Eq.(\ref{estadogen}))
\begin{equation}
 \rho_{\rm r}(0)= \frac{1-r}{4} I + r |\phi\rangle \langle \phi |,
\label{werner}
\end{equation}
where $r~\epsilon(0,1]$
determines the mixing of the state and $I$ is the unit matrix
in the Hilbert space $2\times2$. The state $|\phi\rangle$ may be any of the
following states,
\begin{eqnarray}
 |\vartheta \rangle &=& \sqrt{1-p} |0 0 \rangle + \sqrt{p} |1 1 \rangle
\label{00}\\
|\mu \rangle &=& \sqrt{1-p} |0 1 \rangle + \sqrt{p} |1 0 \rangle
\label{01}
\end{eqnarray}
where $p$ determines the degree of entanglement being
${|0\rangle,|1\rangle}$ eigenstates of the Pauli operator
$\sigma_z$. It is easy to note that when $p=1/2$, Eqs.(\ref{00}) and
(\ref{01}) are Bell states and Eq.(\ref{werner}) defines the
so-called Werner states which play an important role in quantum
information processing. We use
Eq.(\ref{werner}) because it includes all possible cases, such as
pure or mixed states and maximal or non-maximal entangled states.
The first term in Eq.(\ref{werner}) can be regarded as the noise and
the mixing coefficient $r$ describes the intensity of noise.
Recently the one-to-one correspondence between r of the Werner state
and the temperature T of the one-dimensional Heisenberg two-spin
chain with a magnetic field B along the z-axis, has been established
\cite{batle}.

Finally, if we assume an ohmic environment at zero temperature,
the decoherence factors take the following forms,
\begin{eqnarray}
 \Gamma_{1}(t) &=& e^{- 2{\gamma_0}_1 \log ({1 + \Lambda^2 t^2})}, \\
 \Gamma_{2}(t) &=& e^{- 2{\gamma_0}_2 \log({1 + \Lambda^2 t^2})}, \\
 \Gamma_{12}(t) &=& e^{- 2{\gamma_0}_{12} \log({1 + \Lambda^2 t^2})},
\end{eqnarray}
similarly to the spin-boson model for zero temperature \cite{PRA},
where $\Lambda$ is the environmental frequency cutoff.

At this stage, we know the dynamics of the bipartite system for all
times. However, we are interested in the geometric phase adquired by
the composite system in one quasicyclic evolution $\tau \sim
2\pi/\Omega$, being $\Omega$ the characteristic frequency of the
bipartite system. As we have mentioned in the Introduction, the
kinematic approach to the geometric phase can be done by the use of
the reduced density matrix. Therefore, we can compute the
eigenvalues and eigenvectors of the matrix and compute
the geometric phase of this state.\\

\subsection{Werner state with $|\vartheta \rangle$}

We shall start by assuming an initial density matrix of the
form of Eqs.(\ref{werner}) and (\ref{00}). In such a case,
 it is possible to write the reduced density matrix for
all time $t>0$ as

\begin{widetext}
\begin{equation}
\rho_{\rm r_A}(t)=\left(\begin{array}{cccc} \frac{(1-r)}{4} + r(1-p) & 0 & 0
& r \sqrt{p(1-p)}  e^{-{\rm i (\Omega_1 +\Omega_2) t}}
\Gamma_1(t) \Gamma_2(t) \Gamma_{12}^2(t)\\
0 & \frac{(1-r)}{4} & 0 & 0 \\
0 & 0 & \frac{(1-r)}{4} & 0 \\
r \sqrt{p(1-p)}  e^{{\rm i (\Omega_1+\Omega_2) t}} \Gamma_1(t)
\Gamma_2(t) \Gamma_{12}^2(t)  & 0 & 0 & \frac{(1-r)}{4} + r p
\end{array}\right).
\label{rhoconr}
\end{equation}
\end{widetext}

In this case, the eigenvalues  of Eq.(\ref{rhoconr}) are
\begin{eqnarray}
\varepsilon_+(t) &=& \frac{1}{4} \bigg( 1 + r + 2 r \sqrt{1+4 p
(1-p)
(\Gamma_{12}^4 \Gamma_1^2 \Gamma_2^2 -1)} \bigg), \nonumber \\
\varepsilon_-(t) &=& \frac{1}{4} \bigg( 1 + r - 2 r \sqrt{1+4 p
(1-p)
(\Gamma_{12}^4 \Gamma_1^2 \Gamma_2^2 -1)} \bigg), \nonumber \\
\varepsilon_1 &=& \frac{1-r}{4}, \nonumber \\
\varepsilon_2 &=& \frac{1-r}{4}, \nonumber
\end{eqnarray}
with the corresponding eigenvectors,

\begin{eqnarray}
\vert \Psi_+ \rangle &=&  \frac{ r \sqrt{p(1-p)} e^{-{\rm i (\Omega_1 +\Omega_2) t}}
 \Gamma_{12}^2 \Gamma_1 \Gamma_2}  {\sqrt{(
\varepsilon_+ -(1-p))^2+  p(1-p) r^2 \Gamma_1^2 \Gamma_2^2
\Gamma_{12}^4}}
 \vert 0 0 \rangle \nonumber  \\
& +& \frac{(\varepsilon_+ - (1-p))}{\sqrt{(\varepsilon_+ - (1-p))^2
+  p(1-p) r^ 2 \Gamma_1^2 \Gamma_2^2 \Gamma_{12}^4}}
\vert 1 1 \rangle \nonumber \\
\vert \Psi_- \rangle &=&  \frac{ r \sqrt{ p(1-p)}  e^{-{\rm i
(\Omega_1 +\Omega_2) t }} \Gamma_{12}^2 \Gamma_1 \Gamma_2} {\sqrt{(
\varepsilon_- -(1-p))^2+  p(1-p) r^ 2 \Gamma_1^2 \Gamma_2^2
\Gamma_{12}^4}}
\vert 0 0 \rangle \nonumber \\
& +& \frac{(\varepsilon_- - (1-p))}{\sqrt{(\varepsilon_- - (1-p))^2
+  p(1-p) r^2 \Gamma_1^2 \Gamma_2^2 \Gamma_{12}^4}}
\vert 1 1 \rangle \nonumber \\
\vert \Psi_1 \rangle &=& \vert 0 1 \rangle \nonumber \\
\vert \Psi_2 \rangle &=& \vert 1 0 \rangle. \nonumber \\
\end{eqnarray}

We see that, with this particular choice of the
initial state, we are left to work in the space spanned by $\vert 0 0 \rangle$
and $\vert 1 1 \rangle$, since $|\Psi_{1,2} \rangle$
are time independent and hence will not contribute to the
geometric phase. Using Eq.(\ref{fasegeo}),
formally the geometric phase can be
computed as
\begin{eqnarray}
\phi_G &=& \rm {arg}\bigg\{\sqrt{\varepsilon_+(0)\varepsilon_+(\tau)} \langle \Psi_+(0)|
\Psi_+(\tau) \rangle \nonumber \\
& & e^ {i ~{\Omega} \int_0^ {\tau} \cos^ 2(\theta_+(t_1))~dt_1} \nonumber \\
&+& \sqrt{\varepsilon_-(0)\varepsilon_-(\tau)} \langle \Psi_-(0)|
\Psi_-(\tau) \rangle \nonumber \\
&& e^ {i~ {\Omega} \int_0^ {\tau} \cos^
2(\theta_-(t_1))~dt_1}\bigg\}, \label{fasegen}
\end{eqnarray}
by defining
\begin{equation}
 \cos(\theta_\pm(t))= \frac{r \sqrt{p(1-p)} \Gamma_1 \Gamma_2 \Gamma_{12}^2}{
(\sqrt{(\varepsilon_\pm -(1-p))^2+  p(1-p) \Gamma_{12}^4 \Gamma_1^2
\Gamma_2^2})}, 
\label{cosenotheta+-}
\end{equation}
in analogy to what has been done for a single two-level system in
$\rm SU(2)$ \cite{Tong,PRA}. In this case, one can use $\tau=
2\pi/{\Omega}$, with ${\Omega}=(\Omega_1 + \Omega_2)$.\\

In order to compute the geometric phase and obtain concrete results,
we shall choose $r=1$. In this case it is easy to see that
Eq.(\ref{fasegen}) simplifies since $\varepsilon_-(0)=0$ and the
contribution of this eigenenergy and its associated eigenvector to
the geometric phase is null. Hence, the the geometric phase becomes
\begin{eqnarray}
\phi_G &=&  (\Omega_1 +\Omega_2) \times \nonumber \\
& & \int_0^{\tau}  dt~\frac{(1-p)p \Gamma_{12}^4 \Gamma_1^2
\Gamma_2^2} {(1-p)p\Gamma_{12}^4 \Gamma_1^2 \Gamma_2^2
+ [\varepsilon_+ -(1-p)]^2}. \nonumber \\
&=& (\Omega_1 +\Omega_2) \int_0^{\tau} \cos^2(\theta_+(t)).
\label{geometricphaseSBe1}
\end{eqnarray}\\

\begin{figure}[!ht]
\centering
\includegraphics[width=8.1cm]{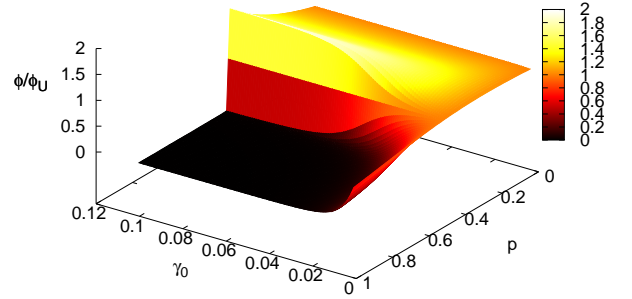}
\caption{Deviation from the unitary geometric phase for the
bipartite system coupled to an ohmic bosonic environment as a
function of the initially entanglement of the state $p$ and the
coupling constant $\gamma_0$. The frequency cutoff is 
$\Lambda /\Omega=100$.} \label{fasevspISBe1}
\end{figure}

It is important to note that if the bipartite system is isolated,
then $\Gamma_i(t)=1$, and the geometric phase becomes
\begin{eqnarray}
\phi_G = (\Omega_1 +\Omega_2) \int_0^{\tau}  dt~(1-p) = 2 \pi (1-p),
\nonumber
\end{eqnarray}
which, of course, is the unitary geometric phase $\phi^U_G$. We can
interpret this result by analogy to a single spin-boson model. In
Ref.\cite{PRA}, we have started by a pure initial state on the Bloch
Sphere, characterized by $\theta_0$, i.e. $| \Phi(0) \rangle =
\cos(\theta_0/2) |0\rangle + \sin(\theta_0/2)|1 \rangle$. Then, in
our present case, it is enough to define
$\sqrt{1-p}=\cos(\theta_0/2)$. Then the unitary geometric phase
becomes $\phi_G^U = 2 \pi \cos^2(\theta_0/2)=\pi (1+\cos\theta_0)$,
which agrees with the geometric phase acquired by a spin in $\rm
SU(2)$. In other words, as the subspace spanned by $\{|0 1 \rangle,
| 1 0 \rangle\}$ is a decoherence free space, it is not affected by
the presence of an environment. Since $[H_S,H_I]=0$, in this case,
it does not evolve in time (see Eq.(\ref{rhorfinal})). Then, as long
as the whole system is prepared in a separable state, the geometric
phase of the composite is exactly that of the evolution happening in
the Hilbert Space spanned by $\{|0 0 \rangle, | 1 1 \rangle \}$,
similarly to that of $2\times 2$.

It is also important to note that in the case of having a maximally
entangled state is $\phi_G = \pi$. Physically, this phase depends on 
the parity of the number of times the state 
crosses the space orthogonal to the initial state in the representation 
of a SO3 sphere. It is already known that for a qubit the total phase gained is 
$\pi$ (or n $\pi$) and it is due to a  
combination of the dynamical phase and the geometric phase. For a two-level 
bipartite system with an arbitrary degree of entanglement a third 
possible type of global phase can be identified (in addition to the dynamical and 
geometric phase already known): a topological phase, which is a consequence of the 
geometry of the entangled two-level system. This phase has been studied for maximally 
entangled states (MES) and it is at the origin of singularities appearing in the phase 
of MES during a cyclic evolution. In \cite{milman}, it is studied the phase
dynamics of entangled qubits under unitary cyclic evolutions. Therein,
it is shown that, after a cyclic evolution, the combination of the different phases 
always leads to a global phase of an entire 
multiple of $\pi$. This result, already known and verified experimentally for a single 
qubit is recovered here for an entangled qubit with maximal degree of entanglement
in the presence of an environment.

For a bipartite state, we can not longer use the Bloch sphere to seek a geometric 
representation of that state. In \cite{milman} it has been shown that a geometric 
representantion of a bipartite state can be obtained by using a Bloch ball and a SO3 sphere. 
Therefore,  the total phase gained by a state is a combination of not only the dynamical 
and geometrical phase, but also the topological phase. Similarly to one qubit states, 
MES also gain a total phase of $\pi$ (or $n \pi$) under a cyclic evolution. However, 
this phase is of topological origin. 
In this context, we can explain our results. We are looking to corrections to the unitary
phase. However, in the case of a MES  there is no correction to the unitary phase. 
How can this be? We strongly believe that it is due to the nature of the
unitary phase. As we have explained above, it is of topological origin. Then, 
it can not be disturbed or
modified by the dynamic of the environment. For all other degree of entanglement, 
we do obtain a correction displaying the type of behaviours stated in \cite{milman}, 
with the interesting additive factor that we are considering the presence of an enviroment.

In Fig.\ref{fasevspISBe1}, we show the behavior of the geometric
phase for the open composite system, as a function of $p$ and the
coupling constant $\gamma_0$, when the bipartite system is coupled
to a bosonic ohmic environment by the same dimensionless coupling,
i.e. ${\gamma_0}_1= {\gamma_0}_2={\gamma_0}_{12}\equiv \gamma_0$.
Therein, we can note that for very small values of the coupling
constant, the geometric phase is that of the unitary
system, meaning by the latter that there is no interaction of the
bipartite with an environment (closed system). When the presence of
the environment is relevant enough to influence the dynamics of the
bipartite system, the correction to the unitary geometric phase
$\delta\phi_G$ also depends of the degree of entanglement $p$, being
zero when the initial state is maximally entangled ($p=1/2$). This
result has some peculiar characteristics similar to the ones
observed in \cite{cui}. There, authors studied the isolated system
and obtained the geometric phase by tracing over one spin as
suggested in \cite{Sjoqvist}. Even though we are considering an
external environment, we can note the symmetry in $p$ of our result
derived from the type of coupling, and also the jump at the
crossover point $p=1/2$ ($\theta_0=\pi/2$), when the initial state
falls from the upper to the lower semi-sphere in the Bloch
representation.

In order to analitically analyze the correction to the unitary geometric
phase we can perform a series expansion in powers of the
 coupling with the environment $\gamma_0$ (assuming
a solely coupling between the bipartite system and
the environment, i.e. ${\gamma_0}_i=\gamma_0$))
\begin{eqnarray}
 \phi_G &\approx & 2 \pi (1-p) + 32 \gamma_0 p
\frac{\Omega}{\Lambda}
(1-3p + 2p^2) \nonumber \\
& &\bigg[ \rm \arctan(2 \pi
\frac{\Lambda}{\Omega})  \nonumber \\
&+& \pi \frac{\Lambda}{\Omega} \bigg(-2
+ \log\bigg(1+4 \pi^2\frac{\Lambda^2}
{\Omega^2}\bigg)\bigg) \bigg],
\label{geopertSBe1}
\end{eqnarray}
where $\Omega = \Omega_1 + \Omega_2$.\\

Assuming that $\Lambda / \Omega\gg 1$, the correction
$\delta\phi_G$ to the geometric phase can be well approximated by

\begin{equation}
\delta\phi_G \approx 64\pi \gamma_0 p(1-3p +
2p^2)\left(\log\left(\frac{2\pi\Lambda}{\Omega}\right) -
1\right)\label{pertur1},
\end{equation}
which is essentially the same correction found in \cite{PRA} for a
single two-level particle.

\begin{figure}[!ht]
\centering
\includegraphics[width=8.1cm]{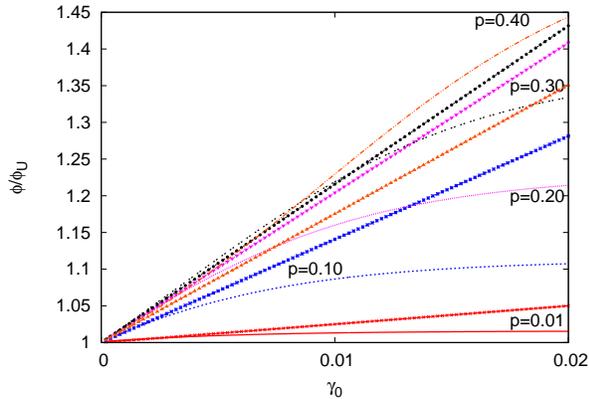}
\caption{(colors online) Deviation from the unitary geometric phase
for the bipartite system coupled to an ohmic bosonic environment as
a function of the initially entanglement of the state $p$ and the
coupling constant $\gamma_0$. Exact results are plotted with lines
and perturbative ones with lines and dots. Equal colors indicate
equal values of $p$. $\Lambda /\Omega =100$.} \label{pertSBe1}
\end{figure}

In Fig.\ref{pertSBe1} we show the geometric phase from
Eq.(\ref{geometricphaseSBe1}) and the geometric phase obtained by
the use of Eq.(\ref{geopertSBe1}). The above expressions show that
the phase of the composite system depends on several parameters: the
coupling constant $\gamma_0$, the frequency cutoff $\Lambda$ (i.e.
the ''size`` of the environment) and the degree of entanglement of
the initial state $p$. In Fig.\ref{pertSBe1}, we can observe that
the correction grows with $p$ up to $p=1/2$. For values of $\gamma_0
p \ll 1$ the perturbative expression for the geometric phase becomes
a good approximation.\\

\begin{figure}[!ht]
\centering
\includegraphics[width=8.1cm]{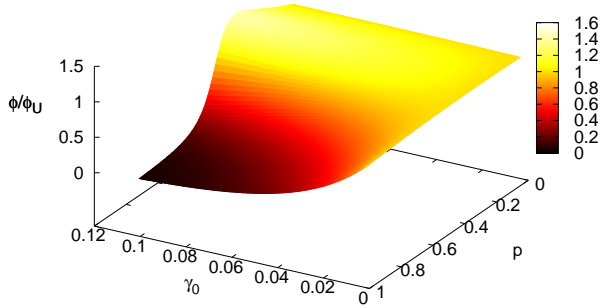}
\caption{(colors online) Deviation from the unitary geometric phase
for the bipartite system coupled to an supraohmic bosonic
environment as a function of the initially entanglement of the state
$p$ and the coupling constant $\gamma_0$. $\Lambda /\Omega=100$.}
\label{fasespIISBe1}
\end{figure}

We can think of a realistic model where our central system is
coupled to an external environment composed of phonons or photons.
In such case, we should consider an non-ohmic spectral density for
such an environment, particularly one that goes as $J(\omega) \sim
\omega^3$, usually called supraohmic environment \cite{Leggett}. The
decoherence factor takes a subtly different form, being decoherence
even less effective at zero temperature,
\begin{equation}
\Gamma^{\rm supra} (t)= e^{-4 \gamma_0 \frac{\Lambda^4 t^4}{(1 + \Lambda^2 t^2)^2}}.
\end{equation}
For times $\Lambda t > 1$, this factor becomes $\Gamma^{\rm supra} (t)\sim 
e^{-4 \gamma_0}$. Since $\gamma_0 < 1$, decoherence is not as effective as 
in previous the example\cite{PRA, PLA}.
In Fig.\ref{fasespIISBe1}, we show the geometric phase for the open bipartite system
when it is coupled to a bosonic supraohmic environment by the same coupling constant
$\gamma_0$. In this case, we see that the effect of the environment over
the geometric phase is less relevant since the correction to the
unitary geometric phase is smaller than the previous case for the same set 
of parameters.

A series expansion in powers of the coupling constant $\gamma_0$, gives
\begin{eqnarray}
\phi_G &\approx &  2 \pi (1-p) + 8 \gamma_0 p \frac{\Omega}{\Lambda}
(1-3p+2p^2) \nonumber \\
&&\bigg[ \pi \Lambda \bigg( \frac{4}{\Omega} +
\frac{2 \Omega}{\Omega^2 + 4 \pi^2 \Lambda^2} \bigg)
- 3 \rm \arctan (2 \pi \frac{\Lambda}{\Omega} )
\bigg], \nonumber \\
\label{geopertSBe1sup}
\end{eqnarray}
where the correction to the phase can also approximated for
$\Lambda/\Omega\gg 1$ as

\begin{equation}
\delta\phi_G \approx 32 \pi \gamma_0 p(1-3p+2p^2),\end{equation} showing it is
smaller than the correction induced in the ohmic case Eq.(\ref{pertur1}) 
by a $\log(\Lambda /\Omega)$ factor \cite{PRA}.
\begin{figure}[!ht]
\centering
\includegraphics[width=8.1cm]{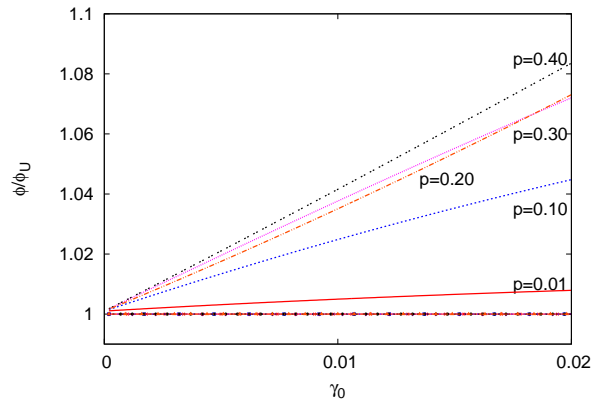}
\caption{(colors online) Deviation from the unitary geometric phase
for the bipartite system coupled to a nonohmic bosonic environment
as a function of the initially entanglement of the state $p$ and the
coupling constant $\gamma_0$. Exact results are plotted with lines
and perturbative ones with lines and dots. Equal colors indicate
equal values of $p$. $\Lambda /\Omega=100$.} \label{pertSBe1nohm}
\end{figure}

In Fig.\ref{pertSBe1nohm} we present the deviation from the unitary
geometric phase for the bipartite system coupled to a supraohmic
bosonic environment as a function of the degree of entanglement of
the initial state $p$ and the coupling constant $\gamma_0$. The 
supraohmic environment is less
effective in inducing decoherence on the system and consequently,
affects less the system's dynamics, but it also produces a phase 
correction on the bipartite state.

The correction to the unitary geometric phase naturally depends on
the size of the environment, set by $\Lambda$. If we consider bigger
or smaller ones, the result will be qualitatively the same
\cite{PLA,PRA}. Typically, decoherence from the ohmic bath is larger,
limiting the possibility of measuring the correction to the phase
using interferometry.

\subsubsection{Linear Entropy and Concurrence}

Entanglement is a quantum-mechanical feature which does
not exist in the classical world. It carries non-local
correlations between the different parts in such a way
that can not be described classically.
Bipartite entanglement of pure states is conceptually
well understood. A useful measure of many-body entanglement
when the total system is in a pure state is the Von Neumann entanglement
entropy $Sl(t)$, which provides a measure of the bipartite
entanglement present in pure states. To be precise, the entanglement
entropy measures the optimal rate at which it is possible to distill
Bell pairs by local operations in the limit of having an
infinite number of copies of the bipartite system.

\begin{figure}[!ht]
\centering
\includegraphics[width=8.1cm]{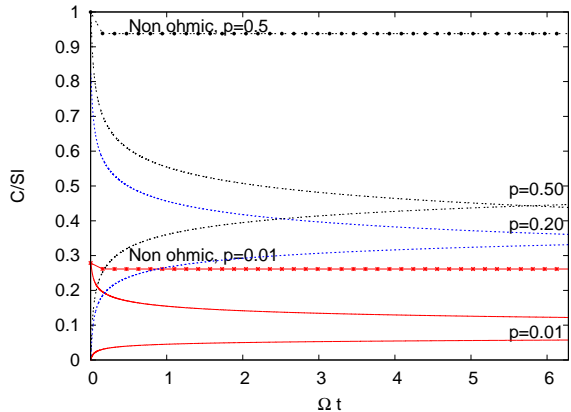}
\caption{(colors online) Concurrence and Linear Entropy as a
function of time for different values of the parameter $p$ when the
bipartite system is coupled to a bosonic environment with coupling
constant $\gamma_0=0.002$.  Parameters used: $\Lambda=100$, $p=0.01$
solid line, $p=0.20$ dashed lined and $p=0.5$ dot-dashed line. The
concurrence for a supraohmic environment (with dots) is also indicated for
$p=0.01$ and $p=0.5$.} \label{concentISBe1}
\end{figure}

The von Neumann entanglement entropy (or linear entropy) is obtained
by focusing on bipartite systems where space can be divided into two
regions: the one corresponding to the bipartite system and the one
of the environment. After tracing over the degrees of freedom of the
environment, we obtain the reduced density matrix of the system
$\rho_{\rm r}$. The von Neumann entanglement entropy is defined as,
\begin{equation}
 Sl= -\rm Tr[\rho_{\rm r} \log_2(\rho_{\rm r})].
\end{equation}
In terms of the eigenvalues of the reduced density matrix of the
bipartite system, the entanglement entropy $Sl$ reads,
\begin{equation}
 Sl(t)= -\varepsilon_+(t) \log_2(\varepsilon_+(t))-
\varepsilon_-(t) \log_2(\varepsilon_-(t)). \label{entSBe1}
\end{equation}
Quantum decoherence implies a rapid reduction of the
off-diagonal terms of the bipartite reduced
density matrix which results in the case of maximal
entanglement in $\varepsilon_{\pm}(t) \rightarrow 1/2 $
and $Sl(t)\rightarrow 1$.\\

The quantity for measuring the entanglement between the different
parts of the composite system is the concurrence \cite{Wooters}. The
concurrence for the evolution of this state can be computed as
${\cal C}(\rho_{c_{\rm r}})= {\rm
max}(0,\sqrt{\lambda_1}-\sqrt{\lambda_2}-
\sqrt{\lambda_3}-\sqrt{\lambda_4})$, where
$\lambda_1,~\lambda_2,~\lambda_3$ and $\lambda_4$ are the
eigenvalues of $\rho_{c_{\rm r}}= \rho_{\rm r}^*(\sigma_y^1 \otimes
\sigma_y^2) \rho_r (\sigma_y^1 \otimes \sigma_y^2)$. 

In the case of the initial condition considered in the last Section, 
the concurrence is
\begin{equation}
 {\cal C}_A= 2 \Gamma^4(t) \sqrt{p(1-p)}.
\label{concSBe1}
\end{equation}

In the closed system case, the concurrence is ${\cal C}_A= 1$ when
considering a MES ($p = 1/2$), and ${\cal C}_A= 0$ for product sates
($p=0$). In the open system case, the concurrence depends on the
decoherence factor $\Gamma(t)$ as can be seen from
Eq.(\ref{concSBe1}), being smaller in the supraohmic case.

\begin{figure}[!ht]
\centering
\includegraphics[width=8.1cm]{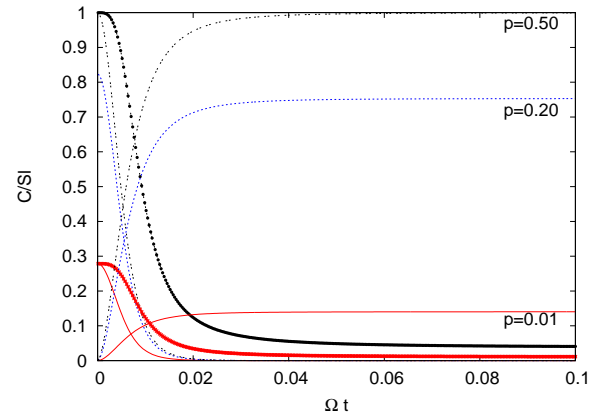}
\caption{(colors online)
Concurrence and Linear Entropy
 as a function of time for different values of the parameter $p$
when the bipartite system is coupled to a bosonic environment with coupling constant
$\gamma_0=0.1$. Parameters used:  $\Lambda=100$, $p=0.01$ solid line, $p=0.20$
dashed lined and $p=0.5$ dot-dashed line.
The concurrence for a supraohmic environment is also indicated (with dots) for
$p=0.01$ and $p=0.5$.}
\label{concentIISBe1}
\end{figure}

In Figs.\ref{concentISBe1} and \ref{concentIISBe1} we show the
linear entropy (Eq.(\ref{entSBe1})) and concurrence
(Eq.(\ref{concSBe1})) when the bipartite system is coupled to the
bosonic environment by a solely coupling constant $\gamma_0$.
Therein, we can see that when the value of $\gamma_0$ is small, then
the ohmic and supraohmic environments induce the same behaviour on
the entanglement of the system state. However, as $\gamma_0$
increases, the ohmic environment destroys whatever degree of
entanglement might be while the supraohmic is not effective in such
a task.

It is worth mentioning that both quantities, i.e the geometric phase
and concurrence, are modified by the presence of the environment to
a greeter or lesser extend. We can see that for a strong environment
($\gamma_0=0.1$), and entangled states between $0<p<1/2$ and
$1/2<p<1$, the correction to the geometric phase is important.
However, for all values of $p$ the concurrence decays at very short
times in the case of the ohmic environment. In the case of the
supraohmic environment, the situation is much similar to a
quasi-isolated system, and the
concurrence is not much affected by the noise and dissipation
introduced by the external bath.

Finally, it has been stated that there was some kind of correlation
among the geometric phase of the individual spins and the 
concurrence for an entangled state of two spin-1/2 particles \cite{basu}. However,
we have shown that there is no such relations in the general case
(when the coupling to an external environment is considered). Only
in the isolated (or unitary) case, it is possible to explicitly show
that
\begin{equation}
\frac{\phi^U_G}{{\cal C}_A} = \pi \sqrt{\frac{1 - p}{p}},
\end{equation}
which is constant for a fixed value of p. In the case of a MES with $p=1/2$, the above
relation reads $ \Phi^U_G =  \pi {\cal C}_A$ \cite{basu}.
\\

\subsection{Werner state with $|\mu \rangle$}

Another possibility is to begin with an initial bipartite state
of the form $|\mu \rangle$ in Eq.(\ref{werner}).
In this case, the reduced density matrix
has a slightly simpler expression such as,
\begin{widetext}
\begin{equation}
\rho_{\rm r_B}(t)=\left(\begin{array}{cccc} \frac{(1-r)}{4}  & 0 &
0 & 0 \\
0 & \frac{(1-r)}{4} + r(1-p) &  r \sqrt{p(1-p)} e^{-i(\Omega_1 - \Omega_2) t}
 \Gamma_1 \Gamma_2 \tilde{\Gamma}_{12}^2  & 0\\
0 &  r \sqrt{p(1-p)} e^{i(\Omega_1 - \Omega_2) t} \Gamma_1
\Gamma_2 \tilde{\Gamma}_{12}^2   & \frac{(1-r)}{4}  + r p & 0 \\
0 & 0 & 0 & \frac{(1-r)}{4}
\end{array}\right). \nonumber
\end{equation}
\end{widetext}

As before, we need to compute the eigenvalues and eigenvectors of
the reduced density matrix in order to know the geometric phase of
the system. For simplicity, we shall choose again $r=1$. In that
case, the eigenvalues will be very similar to those calculated
before, with $\varepsilon_1=\varepsilon_2= 0$ and

\begin{eqnarray}
\varepsilon_+(t) &=& \frac{1}{2} \bigg(1 + \sqrt{1+4 p (1-p)
(\tilde{\Gamma}_{12}^4 \Gamma_1^2 \Gamma_2^2 -1)} \bigg), \nonumber \\
\varepsilon_-(t) &=& \frac{1}{2} \bigg(1 - \sqrt{1+4 p (1-p)
(\tilde{\Gamma}_{12}^4 \Gamma_1^2 \Gamma_2^2 -1)} \bigg), \nonumber
\end{eqnarray}
with the corresponding eigenvectors that can be easily computed as before.

Again, we can see that $\varepsilon_-(t=0)=0$. Then, the only contribution
to the geometric phase comes from the eigenvalue $\varepsilon_+$ and its
associated eigenvector $|\Psi_+ \rangle$.  Similarly to the above procedure,
the geometric phase is,
\begin{eqnarray}
\phi_G &=& (\Omega_1 - \Omega_2) \times \nonumber \\
&& \int_0^{\tau}~dt  \frac{p(1-p) \Gamma_1^2 \Gamma_2^2
\tilde{\Gamma}_{12}^4} {p(1-p) \Gamma_1^2 \Gamma_2^2
\tilde{\Gamma}_{12}^4+ [\varepsilon_+ - (1-p) ]^2}. \nonumber
\end{eqnarray}

From this equation, it is possible to show that there is no
correction to the unitary phase in the case of starting whit a MES.
Therefore, when $p = 1/2$ we get $\phi_G = \pi$.

As expected, we reobtain the unitary result for the geometric phase
if the system is isolated, i.e. ${\gamma_0}_i=0$, and consequently
$\Gamma_i=1$. Furthermore, if we assume that the coupling between
each spin and the environment is the same, i.e.
${\gamma_0}_1={\gamma_0}_2={\gamma_0}_{12}\equiv \gamma_0$, we see
that surprisingly there is no correction to the unitary geometric
phase, since
\begin{equation}
\phi_G = \phi^U_G = 2 \pi (1-p) \nonumber
\end{equation}
being at this time $\tau = 2\pi/{\Omega}$ with
$\Omega=(\Omega_1 - \Omega_2)$.

We can state that in this case the geometric phase is robust
against the action of the external environment. This is easily
understood since, as we have explained before, the subspace spanned
by ${| 0 1\rangle, |1 0 \rangle }$ is decoherence-free. Solely in
the case in which all the coupling strengths are different, one to
each other, the total geometric phase accounts for environmentally
induced corrections.

Incidentally, if  we compute the concurrence for this initial
bipartite system state, we see that it is no longer a function of
the decoherence factor (when ${\gamma_0}_i\equiv \gamma_0$),

\begin{eqnarray}
{\cal C}_B &=& \sqrt{1-p + 2 \sqrt{p(1-p)^3} } \nonumber \\
&-& \sqrt{1-p- 2 \sqrt{p(1-p)^3} } .\nonumber \\
\label{concSBe2}
\end{eqnarray}

In this case the concurrence ${\cal C}_B$ does not depends on the
decoherence factor and, as in the usual case, we have $ {\cal C}_B =
1$ for $p=1/2$ and $ {\cal C}_B = 0$ for $p=0$.

\section{Bipartite Spin-Spin Model}
\label{SS}

Spin environments are typically the appropriate model in the low
temperature regime. In particular, experiments devoted to the
studies of macroscopic quantum coherence and decoherence require
temperatures close to absolute zero in order to operate.
Experimental evidence shows that decoherence is mainly dominated by
interactions with localized modes in this setting, such as
paramagnetic spins, electronic impurities, defects and
 nuclear spins. Each of the localized modes is described by
a finite-dimensional Hilbert space with a finite energy cutoff.
There are numerous examples that show how spin environments
influence the dynamics of a central system by becoming a source of
decoherence.

In this framework, we examine the decoherence induced by
disordered interacting spin baths at finite temperature.
Our choice of the bath is the most simple case for an Ising
chain so as to facilitate an analytical study of the
decoherence and the geometric phase for the model.
We shall consider the bipartite two-level system
coupled to an external spin environment, modelled by the
following Hamiltonians:
\begin{eqnarray}
H_S &=& \frac{\hbar}{2}(\Omega_1 \sigma_z^1 + \Omega_2 \sigma_z^2)
+ \gamma \sigma_z^1 \otimes \sigma_z^2 \nonumber \\
H_I &=& \sigma_z^1 \otimes \sum_{i=1}^N \varepsilon_i \sigma_{z i}
+ \sigma_z^2 \otimes \sum_{i=1}^N \lambda_{i} \sigma_{z i}  \nonumber \\
H_B &=& \sum_{i=1}^N h_i \sigma_{x i}.
\end{eqnarray}

We have included the free Hamiltonian $H_B$ of the environment,
where $h_i$ denotes the tunneling matrix element for the
ith-environmental spin. This free Hamiltonian lends intrinsic
dynamics to the environment, in contrast with the more simplified
spin-environment models.  

As we have done with the bosonic environment, it is imperious to know the
dynamics of the system in order to study the geometric phase
from the kinematic point of view (Eq.(\ref{fasegeo})). To that end, we have
derived the reduced density matrix of the bipartite system in Appendix \ref{derivationSS}.
Herein, we shall consider two different initial states as before (Eqs. (\ref{00}) 
and (\ref{01})).\\

\subsection{Werner state con $|\vartheta \rangle$}

Herein, we shall write
$| \vartheta \rangle$ in Eq.(\ref{werner}).
For this initial state, the reduced density matrix has a much simpler
expression than the general one,

\begin{widetext}
\begin{equation}
\rho_{\rm r_A}(t)=\left(\begin{array}{cccc} \frac{(1-r)}{4} + r(1-p) & 0 & 0
&
r \sqrt{p(1-p)} e^{-i (\Omega_2 + \Omega_1) t} Q(t) \\
0 & \frac{(1-r)}{4} & 0 & 0\\
0 & 0 & \frac{(1-r)}{4} & 0 \\
r \sqrt{p(1-p)} e^{i (\Omega_2 + \Omega_1) t} Q^*(t) &
0& 0 & \frac{(1-r)}{4} + r p
\end{array}\right), \nonumber
\end{equation}
\end{widetext}
where the decoherence factor $Q(t)$ is defined as,
\begin{eqnarray}
 Q(t)&=& \prod_{i=1}^N   \bigg\{ 1 - \bigg( \frac{2 (\varepsilon_i +
\lambda_i)^2}{h_i^2 + (\varepsilon_i + \lambda_i)^2} \bigg) \nonumber
\\
 &\times &\sin^2(t\sqrt{h_i^2+(\varepsilon_i+\lambda_i)^2})
\bigg\} \label{Q}
\end{eqnarray}
and has been derived in the Appendix \ref{derivationSS}.

In order to compute the geometric phase, we need to know the
eigenvalues of the reduced density matrix. By setting $r = 1$, these
are
\begin{eqnarray}
 \epsilon_{1,2} &=& 0 \nonumber \\
\epsilon_{\pm} &=& \frac{1}{2} \bigg[ 1 \pm \sqrt{1 + 4(1-p)p (Q(t)^2-1)}\bigg]
\end{eqnarray}
and the corresponding eigenvectors
\begin{eqnarray}
|v_{1} \rangle &=& |0 1 \rangle \nonumber \\
|v_{2} \rangle &=& |1 0 \rangle \nonumber
\end{eqnarray}
and
\begin{eqnarray}
|v_{+} \rangle &=& \frac{\sqrt{p(1-p)} |Q(t)| e^{-i(\Omega_1+\Omega_2)t}}
{\sqrt{p(1-p) |Q(t)|^2 + [\epsilon_+ -(1-p)]^2}}|0 0 \rangle \nonumber \\
&+& \frac{[\epsilon_+ -(1-p)]}
{\sqrt{p(1-p) |Q(t)|^2 + [\epsilon_+ -(1-p)]^2}}|1 1  \rangle \nonumber \\
|v_{-} \rangle &=& \frac{\sqrt{p(1-p)} |Q(t)| e^{-i(\Omega_1+\Omega_2)t}}
{\sqrt{p(1-p) |Q(t)|^2 + [\epsilon_- -(1-p)]^2}}|0 0 \rangle \nonumber \\
&+& \frac{[\epsilon_- -(1-p)]}
{\sqrt{p(1-p) |Q(t)|^2 + [\epsilon_- -(1-p)]^2}}|1 1  \rangle \nonumber
\end{eqnarray}

\begin{figure}[!ht]
\centering
\includegraphics[width=8.cm]{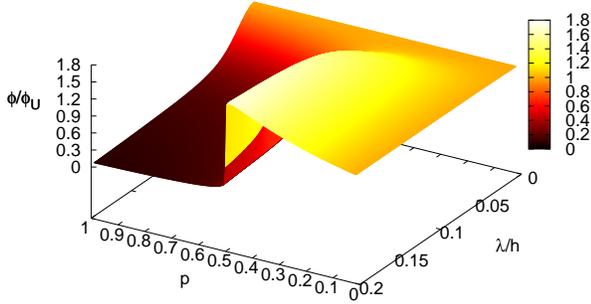}
\caption{(colors online)
Geometric phase for a bipartite system coupled to a spin
environment comprised of $N=100$ spins.}
\label{faseSSe1N10}
\end{figure}

By the use of Eq.(\ref{fasegeo}), the geometric phase becomes,
\begin{equation}
\phi_G = (\Omega_1+\Omega_2) \int_0^{\tau}~dt \frac{p(1-p) |Q(t)|^2}
{p(1-p)|Q(t)|^2 + [\epsilon_+ -(1-p)]^2},
\label{faseSSe1}
\end{equation}
and it is represented in Fig.(\ref{faseSSe1N10}) for an environment
of $N =100$ spins. Therein, we can see that the deviation from the
unitary geometric phase is less than in the ohmic bosonic
environment seen before, but greater than in the supraohmic case.
Then,  if we consider an experiment of a bipartite system coupled to
a spin and photonic environment, we can note that decoherence comes
mainly from the spin environment as state in \cite{Stamp} (assuming
``size-comparable'' environments).

As we have shown for the bosonic environment, the correction to the
geometric phase is zero for an initial MES of the composite system,
getting a total phase of $\pi$.

It is easy to note that we can reobtain the unitary geometric phase
if the bipartite system is isolated. In that case,
$\varepsilon_i=0=\lambda_i$ in $Q(t)$ and the decoherence factor
$Q(t)=1$.  Similarly, we can check that $\epsilon_+=1$ and
$\epsilon_-=0$. Then the geometric phase becomes
\begin{equation}
\phi_G = (\Omega_1+\Omega_2) \int_0^{\tau}~dt (1-p)= \phi_G^U = 2
\pi (1-p),
\end{equation}
assuming $\tau= 2 \pi/(\Omega_1+\Omega_2)$.

We can perform a series expansion in powers of the coupling with the
environment. For that, we shall assume that the couplings between
each spin and the environment are equal, i.e.
$\varepsilon_i=\lambda_i$. In order to achieve an analytical result,
we shall assume all the bath spins to have the same coupling
constant and frequency. Then, the decoherence factor becomes $Q(t)=
\prod_{i=1}^N q_i(t)= q(t)^N$ and we can forget about the product
function. Besides, we shall perform a perturbative expansion in
powers of the dimensionless coupling constant $\lambda/h$. In such a
case,
\begin{eqnarray}
 \phi_G &\approx & 2 \pi (1-p) \nonumber \\
&+& \bigg(\frac{ \lambda}{h} \bigg)^2 16 N p (1-3p + 2p^2) \bigg[ 4
\pi - \frac{\Omega}{h} \sin{\left( 4 \pi \frac{h}{
\Omega}\right)}\bigg], \nonumber \\
\label{pertSSe1}
\end{eqnarray}
where $\Omega=\Omega_1 + \Omega_2$.

\begin{figure}[!ht]
\centering
\includegraphics[width=8.cm]{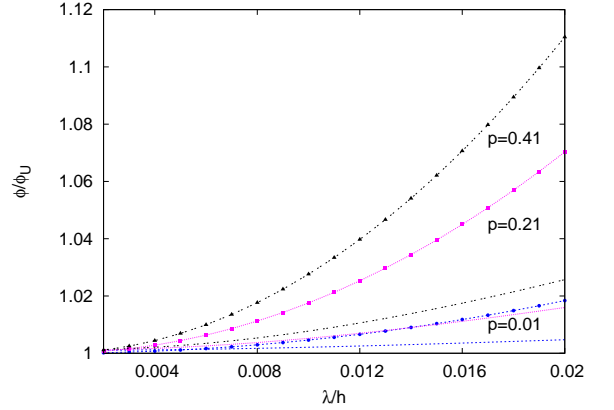}
\caption{(colors online)
Deviation from the unitary geometric phase for the bipartite system 
coupled to an environment comprised of N=100 spins as a function of
 the entanglement of the initial state p and the coupling constant 
$\lambda/h$. Exact results are plotted with lines while  perturbative 
ones with lines and dots. Equal colors indicate equal values of p.}
\label{faseSSe1N102}
\end{figure}

In Fig.\ref{faseSSe1N102} we can see that the series expansion of
the geometric phase results a good approximation of the latter for
mostly all values of $p$. It is more stressed in the case of small
$p N$ since the correction to the phase is quadratic in $\lambda/h$.
\\

\subsubsection{Linear Entropy and Concurrence}

\begin{figure}[!ht]
\centering
\includegraphics[width=8.cm]{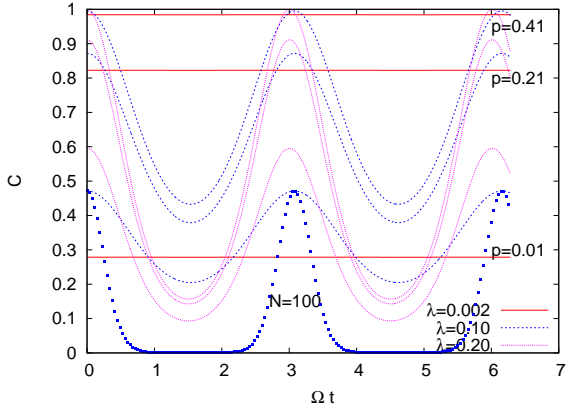}
\caption{(colors online)
Concurrence as a function of time for different values
of the coupling constant
$\lambda$ and different values of the initial state $p$,
when the bipartite system is coupled to
an environment comprised of $N=10$ spins. The dotted line
represents the concurrence for $p=0.01$, $\lambda=0.1$
and $N=100$.}
\label{concvspSSe1}
\end{figure}

As in the other examples analysed for the bosonic environment,
we can study the behaviour
of the concurrence during the dynamics of the bipartite system.
In this case, if we compute the concurrence as explained above,
we obtain
\begin{eqnarray}
 {\cal C}_A &=& \sqrt{p(1-p)(Q(t)+1)^2} -\sqrt{p(1-p)(Q(t)-1)^2 }. \nonumber \\
\label{concSSe1}
\end{eqnarray}

\begin{figure}[!ht]
\centering
\includegraphics[width=8.cm]{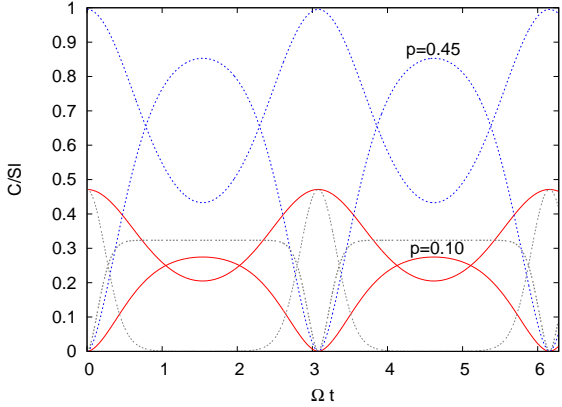}
\caption{(colors online)
Concurrence and Linear Entropy as a function of time
for $\lambda=0.1$ and different values of the initial state $p$,
when the bipartite system is coupled to
an environment comprissed of $N=10$ spins in the case of $p=0.45$.
For $p=0.1$, we present both: an environemt comprised of $N=10$ spins
(red solid line) and $N=100$ spins (grey dotted line).}
\label{concentSSe1}
\end{figure}

In Fig.\ref{concvspSSe1} and \ref{concentSSe1} we plot the
concurrence and linear entropy as a function of the time for
different values of $p$ and the coupling constant
$\lambda=\varepsilon$, for an environment comprised of $N=10$ and
$N=100$ spins. In both Figures, we can see a peculiar behavior for
some of these values. As expected, since we are dealing with a
finite environment, information can be in principle recoverable
since Q(t) is at worst quasiperiodic (both proportional to the
decoherence factor Q(t)). As it is well known for this type of
environment, the effectiveness of the decoherence mechanism is
determined by the dimension of the environment. That is why we
observe the periodic behaviour in the analyzed quantities. On the
other size, when the coupling with the environment is very small,
its presence is not of great importance in the dynamics of the
bipartite, and the concurrence is then constant. However, as we
increase the coupling constant $\lambda$, we find that the
concurrence oscillates due to the oscillating function present in
the decoherence factor. In Fig.\ref{concvspSSe1} we have plotted the
concurrence for different values of $p$ (degree of entanglement of
the initial state) with the same color for equal values of the
coupling constant $\lambda$. All values considered with lines are
for an environment of $N=10$ spins, while the dots represent the
concurrence for $p=0.01$ and $\lambda=0.1$ of an environment of
$N=100$ spins. We can observe that the effect is stronger when the
environment is bigger. Just for the sake of completeness, we plot in
Fig.\ref{concentSSe1} the concurrence and linear entropy for two
extreme values of the degree of entanglement $p$ for a coupling
constant $\lambda=0.1$. In the case of $p=0.1$ we can again observe
the difference in both quantities when the size of the environment
is $N=10$ (solid line) and when it is $N=100$ (grey dotted line).

\subsection{Werner state with $| \mu \rangle$}

Herein, we shall write $| \mu \rangle$ in Eq.(\ref{werner}).
For this initial state, the reduced density matrix has a much simpler
expression,

\begin{widetext}
\begin{equation}
\rho_{\rm rB}(t)=\left(\begin{array}{cccc} \frac{(1-r)}{4} & 0 &
0 & 0 \\
0 & \frac{(1-r)}{4} + r (1-p) & r \sqrt{p(1-p)} e^{-i (\Omega_1 + \Omega_2) t} P(t) & 0\\
0 &  r \sqrt{p(1-p)} e^{i (\Omega_1 + \Omega_2) t} P^*(t)& \frac{(1-r)}{4} + r p & 0 \\
0 & 0 & 0 & \frac{(1-r)}{4}
\end{array}\right), \nonumber
\end{equation}
\end{widetext}
where the decoherence factor $P(t)$ is
\begin{eqnarray}
P(t) &=& \prod_{i=1}^N   \bigg\{ 1 - \bigg( \frac{2 (\varepsilon_i -
\lambda_i)^2}{h_i^2 + (\varepsilon_i - \lambda_i)^2} \bigg)
\nonumber \\
&\times & \sin^2(t\sqrt{h_i^2+(\varepsilon_i-\lambda_i)^2}) \bigg\}. 
\end{eqnarray}
has been defined in the Appendix \ref{derivationSS}.

In order to compute the geometric phase, we need to know the
eigenvalues of the reduced density matrix. These are (for $r = 1$)
\begin{eqnarray}
 \epsilon_{1,2} &=& 0 \nonumber \\
\epsilon_{\pm} &=& \frac{1}{2} \bigg[ 1 \pm \sqrt{1 + 4p (1-p)
(|P(t)|^2-1)}\bigg]
\end{eqnarray}
and the corresponding eigenvectors that again can be easily computed.

By the use of Eq.(\ref{fasegeo}), the geometric phase becomes,
\begin{equation}
\phi_G = \Omega \int_0^{\tau}~dt \frac{p(1-p)|P(t)|^2}{p(1-p)
|P(t)|^2 + [\epsilon_+ -(1-p)]^2},
\nonumber
\end{equation}
where $\Omega=(\Omega_1-\Omega_2)$. 
It is
easy to note that we can reobtain the unitary geometric phase if the
bipartite system is isolated or if it is in a MES with $p = 1/2$. In
isolated from the environment case, $\varepsilon_i=0=\lambda_i$ in
$P(t)$ and the decoherence factor $P(t)=1$.  Similarly, we can check
that $\epsilon_+=1$ and $\epsilon_-=0$. Then the geometric phase
becomes
\begin{equation}
\phi_G = \Omega  \int_0^{\tau}~dt (1-p)= \phi_G^U = 2 \pi
(1-p),
\end{equation}
assuming $\tau= 2 \pi/ \Omega$, once again.

The concurrence for this case can be also calculated as
\begin{eqnarray}
 {\cal C}_B = \sqrt{p(1-p)(P(t)+1)^2} -\sqrt{p(1-p)(P(t)-1)^2 }.\nonumber
\end{eqnarray}

As it is easy to note, this case is similar to the one presented
before but changing the decoherence factor $Q(t)\rightarrow P(t)$.
However this fact is not trivial. If we consider that the coupling between
each spin of the system and the external reservoir is equal,
i.e. $\varepsilon_i=\lambda_i$,
the $P(t)=1$ and the subspace generated by the states $|01\rangle$ and
$|10\rangle$ becomes a decoherence free subspace, reobtaining the
unitary geometric phase and a concurrence ${\cal C}_B = 2 \sqrt{p(1-p)}$.\\

\section{Conclusions}

The geometric phase of entangled states is an issue worth of attention.
It could be a potential application in holonomic quantum computation
since the study of entangled spin systems effectively allows us to contemplate
the design of a solid state quantum computer. However, decoherence is the main
obstacle to overcome. All realistic quantum systems are coupled to their
surroundings to a greater or lesser extent.

We have thoroughly studied the geometric phase for a bipartite
system, i.e. two coupled qubits, also coupled to an external
environment. We have considered both cases: a bosonic environment at
zero temperature (whether ohmic or supraohmic) and a spin
environment. In all cases, we have chosen a general initial state
and computed the evolution of the composite system in terms of the
reduced density matrix.
We have further estimated the decoherence
factors for all cases in order to have a full insight into the
decoherence process induced by the environment on the system's
dynamics. We have seen that there is a hierarchy
among the environments. That means, when considering real
systems, such as single molecule magnets (for example Fe$_8$ molecule),
decoherence induced by the nonlocalized modes of a supraohmic environment
is less than that induced by the localized dicrete modes of a spin environment.

By the use of a kinematic approach, we have computed the geometric
phase $\phi_G = \phi^U_G + \delta \phi_G$ for different choices of
the initial state of the bipartite system. In some cases, we have
also performed a perturbative expansion of the geometric phase in
order to complete the analysis. In all cases, we have found the same
geometric dependence upon the initial angle $\theta_0$ for the
correction of the geometric phase that has been found by some
authors by the use of different approaches \cite{Whitney}.

As entanglement is considered to be one of the key resources in
quantum information science, we have also studied this property of
the composite system. In particular, we have checked that the
correction to the geometric phase is $\delta\phi_G = 0$ for the case of
maximally entangled initial states of the bipartite systems. This is
the case for all the cases considered, no matter the kind of
environment would be present so far.

We have also computed the concurrence and linear entropy of each
initial state. We find a steady relation between the geometric phase
and the concurrence only for a Bell state in the isolated situation.
In all other cases, not only the geometric phase but also the
concurrence are modified but the presence of the environment.

In all cases it is posible to compute the geometric phase in not
only one cyclic but many, i.e. $\tau= n 2 \pi/\Omega$,  with $n$
integer. In that case, the correction to the unitary geometric phase
would be proportional to $n$, the winding number \cite{pau,Whitney}.
This proportionality of $\delta \phi_G$ to $n$ might imply that the
geometric phase of the composite system still has some of its
geometric character. However, this correction is also a function of
the environments spectrum and then the total geometric phase is
not a simple geometric quantity.

\appendix

\section{Derivation of the reduced density matrix for a bosonic environment}
\label{derivationSB}

Herein, we shall derive the reduced density matrix for a two spin $1/2$ 
particles coupled to a bosonic environment through coupling constants
 $\lambda_n$ for the spin 1 and $g_n$ for spin 2 
(being  $\lambda_n$ not necessarily equal to $g_n$).

In order to know the dynamics of the bipartite system, we shall find
the reduced density matrix for the composite system, as has been
done in
\cite{leshouches} for one particle spin-boson model in the case of weak coupling.
We must note that in this case the interaction is
$\tilde{\rm{V}}(t_1)= \sigma_z^1 \sum_n \lambda_n q_n(t_1) +
\sigma_z^2 \sum_n g_n q_n(t_1)$. Similarly to the single spin-boson
model, it is straightforward to compute the time derivative of the
reduced density matrix as (we have set $\hbar = 1$ from here up to
what follows):
\begin{eqnarray}
\dot{\tilde{\rho}}_{\rm r} = - \int_0^t dt_1 {\rm{Tr_B}} \bigg[
\tilde{\rm V}(t_1) , \bigg[ \tilde{\rm{V}}(t_1) , \tilde{\rho}_r
\otimes \tilde{\rho}_B(0)
 \bigg]\bigg]. \nonumber
\end{eqnarray}
The tilde indicates that we are working in the
Interaction Picture.

After doing some algebra, the master equation for the reduced
density matrix can be written as
\begin{eqnarray}
 \dot{{\rho_{\rm r}}}(t) &=& -i [H_s, {\rho_r}] \nonumber \\
&-&\int_0^t dt_1 \bigg\{ \nu_1(t_1)[\sigma_z^1,[\sigma_z^1,{\rho_r}]
+
\nu_2(t_1)[\sigma_z^2,[\sigma_z^2,{\rho_r}]] \nonumber  \\
&+& \nu_{12}(t_1) \bigg([\sigma_z^1,[\sigma_z^2,{\rho_r}]] +
[\sigma_z^2,[\sigma_z^1,{\rho_r}]] \bigg) \bigg\}  \nonumber \\
&+& i  \int_0^t dt_1 ~
\eta_{12}(t_1)\bigg([\sigma_z^1,\{\sigma_z^2,{\rho_r}\}]  \nonumber \\
&+&
[\sigma_z^2,\{\sigma_z^1,{\rho_r}\}]\bigg),
\label{rhorfinal}
\end{eqnarray}\\
where we have already taken the continuum limit and defined the
noise and dissipation kernels as:

\begin{eqnarray}
\nu_i(t)&=&  \int_0^{\infty} d\omega J_i(\omega)
\cos(\omega t) \rm coth(\frac{\beta \omega}{2}), \nonumber \\
\eta_i(t)&=&   \int_0^{\infty} d\omega J_i(\omega)
\sin(\omega t), \nonumber 
\end{eqnarray}
with $i=1,2$ or $i=12$. $J_i(\omega)$ is the spectral density of the environment
  associated to each spin of the system, and $\nu_i(t)$ the corresponding
noise kernel while $\eta_i(t)$ is the dissipation kernel associated to 
$J_i(\omega)$.
 One assumption we shall make 
is that $J(\omega)$ is a reasonably smooth function of $\omega$, and
that is of the form $\omega^n$ up to some frequency $\Lambda$ that
may be large compared to $\Omega_1$ and $\Omega_2$. The spectral
density function can be  written as $J_i(\omega)= {\gamma_0}_i/4
~\omega^n \Lambda^{n-1} e^{-\omega/\Lambda} $ \cite{Leggett}. Notice
that at this stage, the coupling constants of the model $\lambda_n$
and $g_n$ have been absorbed in the continuous limit and that
information is now contained in the dimensionless dissipative
constants ${\gamma_0}_1$, ${\gamma_0}_2$, and ${\gamma_0}_{12}$
respectively, defined in the spectral density of the bath.

In Eq.(\ref{rhorfinal}) we note the first difference with the
dephasing single spin-boson \cite{leshouches,PRA}. In the case of
only one spin coupled to an external environment, there is no
dissipation on the main system induced by the environment (it is
just a pure dephasing model). However, in the bipartite system we
see that dissipation appears dispite of the similar coupling between
the system and the environment.

In this context, we can define the decoherence factors $\Gamma_i(t)$ as
\begin{equation}
 \Gamma_i(t)= e^{-4 \int_0^t dt_1 F_i(t_1)},
\end{equation} with
 \begin{eqnarray}
 F_i(t)&=&\int_0^t dt' \nu_i(t') \nonumber
\end{eqnarray}
and the dissipation induced by the environment
as
\begin{eqnarray}
 G_{12}(t)&=&\int_0^t dt' \eta_{12}(t').\nonumber
\end{eqnarray}

\section{Derivation of the reduced density matrix for an environment comprised of spins}
\label{derivationSS}

Herein, we shall derive the reduced density matrix for a bipartite
system coupled to an environment comprised of $N$ spins.
The system's, environment's and interaction hamiltonians have been considered
in the main text, Sec.\ref{SS}.

Let's take an initial separable state of
the complete system, i.e. $\vert \Psi (0) \rangle= \vert \Phi (0)
\rangle \otimes \vert \chi(0) \rangle$, where $\vert \Phi(0)
\rangle$ is the initial state of the bipartite system while $\vert
\chi (0) \rangle$ is for the spin environment. We can note that
$[H_S, H_I]=0$, so we can write the evolution of the complete state
as
\begin{eqnarray}
\vert \Psi(t) \rangle &=& e^{-i (H_S + H_I + H_B) t } \vert
\Psi(0) \rangle  \\
 &=& e^{-i(H_S + H_I)t} \vert \Phi (0)\rangle  \otimes  e^{-i(H_B)t}
\vert  \chi(0) \rangle .\nonumber
\end{eqnarray}

After cumbersome calculations (that we do not consider necessary to
be presented here), we have obtained the following reduced density matrix
for a general initial state $\vert \Phi(0) \rangle= \alpha \vert 0 0
\rangle + \beta \vert 0 1 \rangle + \zeta |1 0 \rangle + \delta | 1
1 \rangle$ of the bipartite system

\begin{widetext}
\begin{equation}
\rho_{\rm r}(t)=\left(\begin{array}{cccc} |\alpha|^2 & \alpha
\beta^* e^{-i (\Omega_2 + 2 \gamma) t} M(t) & \alpha \zeta^* e^{-i
(\Omega_1 + 2 \gamma) t} N(t) &
\alpha \delta^* e^{-i (\Omega_2 + \Omega_1) t} Q(t) \\
\alpha^* \beta e^{i (\Omega_2 + 2 \gamma) t} M^*(t) &
|\beta|^2 & \beta \zeta^* e^{-i(\Omega_1 -\Omega_2)t}
 P(t) & \beta \delta^* e^{-i(\Omega_1 - 2 \gamma)t} R(t)\\
\alpha^* \zeta e^{i (\Omega_1 + 2 \gamma) t} N^*(t) & \zeta \beta^*
e^{i(\Omega_1-\Omega_2)t} P^*(t) & |\zeta|^2 &
\zeta \delta^* e^{-i(\Omega_2 - 2 \gamma) t} S(t) \\
\alpha^* \delta e^{i (\Omega_2 + \Omega_1) t} Q^*(t)  &
\delta \beta^* e^{i(\Omega_1 - 2 \gamma)t} R^*(t)& \delta
\zeta^* e^{i(\Omega_2 - 2 \gamma) t} S^*(t) & |\delta|^2
\end{array}\right), \nonumber
\end{equation}
\end{widetext}

In the above expression, the trace over the degrees of freedom of
the spin environment is contained in the decoherence factors $M(t),
~N(t), ~P(t),~Q(t), ~R(t)$, and $S(t)$. These factors have several
expressions. $M(t) = {\rm Tr}_B [ e^{-i(H_B + V_s) t} \rho_B(0)
e^{i(H_B + V_r)t}]$, with $V_s=(\varepsilon_i + \lambda_i)$ and
$V_r=(\varepsilon_i - \lambda_i)$. Similarly, $N(t) = {\rm Tr}_B [
e^{-i(H_B + V_s) t} \rho_B(0) e^{i(H_B - V_r)t}]$, $S(t) = {\rm
Tr}_B [ e^{-i(H_B + V_r) t} \rho_B(0) e^{i(H_B - V_s)t}]$, $R(t) =
{\rm Tr}_B [ e^{-i(H_B + V_r) t} \rho_B(0) e^{i(H_B - V_s)t}]$,
$Q(t) = {\rm Tr}_B [ e^{-i(H_B + V_s) t} \rho_B(0) e^{i(H_B -
V_s)t}]$ and $P(t) = {\rm Tr}_B [ e^{-i(H_B + V_r) t} \rho_B(0)
e^{i(H_B - V_r)t}]$. In this paper, we shall only be interested in
the last two coefficients. Since we can choose a pure state for the
initial state of the bath as $\vert \chi(0) \rangle=\prod_{i=1}^N
(\alpha_i \vert 0_i \rangle + \beta_i \vert 1_i \rangle)$, then
$\rho_B(0)= \vert \chi(0) \rangle \langle \chi(0) \vert$. In order
to have closed expressions for these coefficients, we have finally
assumed $|\alpha_i|=|\beta_i|$ and obtained the following
decoherence factors

\begin{eqnarray}
Q(t)&=& \prod_{i=1}^N   \bigg\{ 1 - \bigg( \frac{2 (\varepsilon_i +
\lambda_i)^2}{h_i^2 + (\varepsilon_i + \lambda_i)^2} \bigg) \nonumber
\\
 &\times &\sin^2(t\sqrt{h_i^2+(\varepsilon_i+\lambda_i)^2})
\bigg\} \nonumber \\
P(t) &=& \prod_{i=1}^N   \bigg\{ 1 - \bigg( \frac{2 (\varepsilon_i -
\lambda_i)^2}{h_i^2 + (\varepsilon_i - \lambda_i)^2} \bigg)
\nonumber \\
&\times & \sin^2(t\sqrt{h_i^2+(\varepsilon_i-\lambda_i)^2}) \bigg\}.
\nonumber
\end{eqnarray}

We can find some common features between these expressions and
 preexisting Literature. For example,
we can note that the factor $P(t)$ that affects the states $|0 1\rangle$
and $|1 0 \rangle$ is similar to that found in \cite{pau} for a two-level
 system coupled to an external environment. In that same case, if we do not
consider the self interaction of the bath, i.e. $h_i=0$, then we get
$P(t)= \prod_{i=1}^N \cos[2 (\varepsilon_i +  \lambda_i) t]$ as found in
\cite{nos,Zurek}. It is also possible
to check that $P(0)=1$. As it was done in \cite{Zurek},
we can evaluate the mean value of the decoherence factor. As in that case,
 $\langle P(t) \rangle_{T \rightarrow \infty}
 \rightarrow 0$. If we
estimate the average dispersion as $\sigma^2 = \langle P(t)^2
\rangle - \langle P(t) \rangle ^2= \sum_{i=1}^N p_i$, with $p_i=(1-
\frac{(\varepsilon_i +\lambda_i)^2} {4 (h_i^2 + (\varepsilon_i +
\lambda_i)^2)})$. Under the assumption that all $p_i$ are
approximately equal, the average fluctuations from zero are \beq
\sigma \sim \frac{1}{\sqrt{N}}. \eeq Therefore, large environments
can effectively induce decoherence on the central spin system.  We
can do the same with all the decoherence factors but in the
following we shall only be interested in factors $P(t)$ and $Q(t)$.

{\bf Acknowledgments.} This work was supported by UBA, CONICET,
and ANPCyT, Argentina.


\begin{thebibliography}{99}
\bibitem{Vidal} G. Vidal, Phys. Rev. Lett. {\bf 91}, 147902 (2003); Phys. Rev. Lett. {\bf 93},
 040502, (2004).
\bibitem{Stamp} Philip C.E. Stamp and Alejandro Gaita-Ari\~no, J. Mat. Chem. {\bf 19}, 1718 (2009).
\bibitem{Leggett} A.J. Leggett, S. Chakravarty, A.T. Dorsey, M.P.A. Fisher,
A. Garg, and W. Zwerger, Rev. Mod. Phys. {\bf 59}, 1 (1987).
\bibitem{Zurek} W.H. Zurek, Phys. Rev. {\bf D 26}, 1862 (1982).
\bibitem{schlosshauer} Maximilian Schlosshauer, {\it Decoherence and the Quantum-To-Classical
Transition}, Springer (2007).
\bibitem{Pancharatman}  S. Pancharatnam, Proc. Indian Acad. Sci. A 44, 247
(1956).
\bibitem{Berry} M.V. Berry, {\it Proc. R. Soc. Lond. A} {\bf 392}, 45 (1984).
\bibitem{Anandan}Y. Aharonov, J. Anandan, Phys. Rev. Lett. {\bf 58},
1593 (1988); J. Anandan, Y. Aharonov, Phys. Rev. {\bf D 38}, 1863
(1988).
\bibitem{Tong} D. M. Tong, E. Sjoqvist, L. C. Kwek,
and C. H. Oh,  Phys. Rev. Lett. {\bf 93}, 080405 (2004); see also
Phys. Rev. Lett. {\bf 95}, 249902 (2005).
\bibitem{PRA} F.C. Lombardo and P.I. Villar, Phys. Rev. A {\bf 74}, 042311 (2006).
\bibitem{nos}F.C. Lombardo and P.I. Villar, Int. J. of Quantum
Information {\bf 6},  707713 (2008).
\bibitem{pau} Paula I. Villar, Phys. Lett. {\bf A} 373, 206 (2009).
\bibitem{Whitney}R.S. Whitney and Y. Gefen,  Phys. Rev. Lett. {\bf 90},
 190402, (2003); R.S. Whitney, Y. Makhlin, A. Shnirman,
 and Y. Gefen,  Phys. Rev. Lett. {\bf 94},
070407 (2005).
\bibitem{Carollo}A. Carollo,  I. Fuentes-Guridi, M. Franca Santos, and V. Vedral,
  Phys. Rev. Lett. {\bf 90}, 160402 (2003);   Phys. Rev. Lett. {\bf 92}, 020402 (2004).
\bibitem{dechiara} G. De Chiara, A. Lozinski, G. M. Palma,
Eur. Phys. J. {\bf D 41}, 179-183 (2007); G. De Chiara, G. M. Palma,
Phys. Rev. Lett. {\bf 91}, 090404 (2003).
\bibitem{Benarjee} S. Benerjee and R. Skikanth, Eur. Phys. J. {\bf D 46}, 335 (2008).
\bibitem{Rezakhani} A.T. Rezakhani and P. Zanardi, Phys. Rev. {\bf A 73}, 012107 (2006).
\bibitem{milman} P\'erola Milman and R\'emy Mosseri, Phys. Rev.
Lett. {\bf 90}, 230403 (2003); P\'erola Milman, Phys. Rev. A {\bf
73}, 062118 (2006).
\bibitem{batle} J. Batle, M. Casas, A. Plastino, A. R. Plastino, Phys. Lett. A {\bf 343}, 12 (2005).
\bibitem{cui}H. T. Cui, L.C. Wang, and X.X.Yi, Eur. Phys. J. D {\bf
41}, 385 (2007).
\bibitem{Sjoqvist} E. Sjoqvist, Arun K. Pati, Artur Ekert, Jeeva S. Anandan, Marie Ericsson, Daniel K. 
L. Oi, and Vlatko Vedral, Phys. Rev . Lett. {\bf 85}, 2845 (2000).
\bibitem{PLA} F.C. Lombardo and P.I. Villar, Phys. Lett. A {\bf
371}, 190 (2007).
\bibitem{Wooters} W. K. Wootters, Phys. Rev. Lett. 80, 2245(1998).
\bibitem{basu} B. Basu, Europhys. Lett. {\bf 73}, 833 (2006).
\bibitem{leshouches} J. P. Paz and W. H. Zurek,
{\it Environment induced superselection and the transition from
quantum to classical} in {\it Coherent matter waves,  Les Houches
Session LXXII}, edited by
 R. Kaiser, C. Westbrook and F. David, EDP Sciences, Springer Verlag
 (Berlin) (2001) 533-614; W.H. Zurek, Rev. Mod. Phys. {\bf 75}, 715 (2003).

\end{thebibliography}
\end{document}